\documentclass[11pt]{article}
\usepackage{amsmath,amsfonts,amssymb,latexsym}

\def\be{\begin{equation}}
\def\ee{\end{equation}}
\def\bq{\begin{eqnarray}}
\def\eq{\end{eqnarray}}
\def\beq{\begin{eqnarray*}}
\def\eeq{\end{eqnarray*}}

\begin{document}
\title{{\Huge Limits of isotropic universes with interacting fluids}}
\author{{\Large Spiros Cotsakis\footnote{\texttt{email:\thinspace
skot@aegean.gr}}\,\,\,
 and Georgia Kittou\footnote{\texttt{email:\thinspace gkittou@aegean.gr}}} \\
Research Group of Geometry, Dynamical Systems and Cosmology \\
University of the Aegean\\ Karlovassi 83 200, Samos, Greece}
\maketitle

\begin{abstract}
\noindent We find various asymptotic limits of universes containing two interacting fluids which interact and exchange energy.  We study the finite-time singularities that may develop in these cosmologies and obtain the asymptotic behaviours of the solutions near such blow up regimes.
\end{abstract}

\noindent In Ref. \cite{kittou1}, we studied the dynamical problem defined in the geometric setup of a flat FRW universe with scale factor $a(t)$ containing two fluids with equations of state
\begin{equation}p_1=(\Gamma-1)\rho_1,\quad p_2=(\gamma -1)\rho_2.
\end{equation}
In particular, we are interested in the behaviour of this system near finite-time singularities, in an effort to unravel the dynamical nature of blow up singularities and other asymptotic solutions that such a model can sustain asymptotically.
The total energy-momentum tensor is given by
$
T^{\mathrm{\,(total)}}_{\mu\nu}=T^{\mathrm{\,(1)}}_{\mu\nu}+T^{\mathrm{\,(2)}}_{\mu\nu},
$
and we consider an interaction of the form
\begin{equation}
s\equiv \frac{H}{a^2}v^0=-\beta H^m\rho_{1}^{\lambda} +\alpha H^n\rho_2^{\,\mu},
\end{equation}
where $\alpha,\beta,m,n,\lambda,\mu$ are constants, $H$ is the expansion rate and $v^0$ is the timelike direction of the fluid velocity.
 The evolution of this system is governed  by the
equations
\begin{eqnarray}\label{sys}
3H^2&=&\rho_1+\rho_2\nonumber\\
\dot{\rho}_1+3H\Gamma\rho_1&=&-\beta H^m\rho_1^{\lambda} +\alpha H^n\rho_{2}^{\,\mu}\\
\dot{\rho}_2+3H\gamma\rho_2&=&\beta H^m\rho_1^{\lambda} -\alpha H^n\rho_{2}^{\,\mu}.\nonumber
\end{eqnarray}
We restrict here to the case where all the exponents  $m,n,\lambda,\mu=1$, the general case being currently under investigation (Ref. \cite{kittou2}). Renaming $x=H$, the Einstein equations  (\ref{sys})
become equivalent to the following cubic dynamical system,
\begin{eqnarray}
\dot x&=&y\\
\dot y&=&-Axy-Bx^3,
\end{eqnarray}
where $A=\alpha+\beta+3\gamma+3\Gamma$,
$B=3(\alpha\Gamma+\beta\gamma+3\Gamma\gamma )/2.$
The possible
solutions of this system `at infinity' correspond to the asymptotic orbits of the vector field
\begin{equation}\label{vf} f(x,y)=(y,-Axy-Bx^3).
\end{equation}
The method of asymptotic
splittings developed in Ref. \cite{ba-co} scrutinizes all possible modes that the vector field
(\ref{vf}) attains on approach to the finite-time singularity
located at $t=0$. These modes correspond to the inequivalent  ways that
(\ref{vf}) splits as $t\rightarrow 0$, and are given by the
following three distinct \emph{decompositions},
\begin{eqnarray}\label{decs1}
f^{(1)}&=&(y,-Axy-Bx^3),\quad\textrm{(all-terms-dominant case)}\\
f^{(2)}&=&(y,-Axy),\\\label{decs3} f^{(3)}&=&(y,-Bx^3).
\end{eqnarray}
When applying  the method of asymptotic splittings in our system, we find that we can have in total seven different families of  asymptotic solutions, cf. Ref. \cite{kittou1}, depending on  the Barrow-Clifton parameter of Ref. \cite{ba-cl1} given by $\delta\equiv B/A^2$. In Ref \cite{kittou1}, we looked for Fuchsian type  solutions, that is when the leading order of the formal asymptotic series is a power-law with rational exponents (in particular there is no constant leading term).
In the limit $\delta\rightarrow 0$, the second decomposition
$f^{(2)}=(y,-Axy)$ leads to the unique balance,
 \begin{equation}
\mathcal{B}^{(2)}_{1}=\left[\mathbf{\Xi},\mathbf{p}\right]= \left[
(2/A,-2/A),(-1,-2)\right] ,\quad A\neq 0, \end{equation}
and this in turn is used to build the self-consistent,
 general (two arbitrary constants, the position of the singularity and the shown coefficient of the asymptotic series) asymptotic Fuchsian solution valid around
the singularity,
\begin{equation} \label{pl1}
x=\frac{2}{A}t^{-1}+c_{21}t-\frac{A}{10}c_{21}^2t^3\cdots ,
\end{equation}
while the $y$ expansion is obtained from the above by
differentiation.

In the limit $\delta\rightarrow -\infty$, the third decomposition of the system has two balances that lead to analogous results, namely,
\begin{equation}
\mathcal{B}^{(3)}_{1,2}=[(\pm\sqrt{2/-B},\mp\sqrt{2/-B}),(-1,-2)], \quad B< 0.
\end{equation}
In this  limit of the parameter $\delta$, we meet a case of `phantom fluids'. More precisely, this asymptotic solution exists provided the fluid parameters $\Gamma, \gamma$ satisfy the condition: $\Gamma\gamma<0$. The general series expansion is then  given by the form
\begin{equation}\label{pl2}x=\pm\sqrt{\frac{-2}{B}}t^{-1}+c_{41}t^3\mp \frac{B}{12}c_{41}^2\sqrt{\frac{-2}{B}}t^7+\cdots . \end{equation}
For the range of parameter values $0<\delta\leq1/8$ (this is the case of `standard decay' of Ref. \cite{kittou1}), the fluids decay in proportion to their energy densities. The `all-terms-dominant' case sustains  the  balances \begin{equation}
\mathcal{B}^{(1)}_{1,2}=\left[\left(\frac{A\pm\sqrt{A^2+8B}}{2B},\frac{-A\mp A\sqrt{A^2+8B}}{2B}\right), (-1,-2)\right],
\end{equation}
and these lead to the  the asymptotic forms
\begin{eqnarray}\label{pl3}
x&=&\frac{3}{A}t^{-1}+c_{11}+\frac{A}{3}c^2_{11}t+\cdots,\quad\text{for }\mathcal{B}^{(1)}_2\text{ and }\delta=1/9,\\
\label{pl4}x&=&\frac{4}{A}t^{-1}, \text{ for } \delta=1/8.
\end{eqnarray}
The most dominant term of the  asymptotic solutions (\ref{pl1})-(\ref{pl4}) on approach to the singularity describes either a collapse (big bang type),  or a big rip singularity.

There are also other families of solutions.
The first decomposition of the system reveals a very complicated singularity for the case $\delta>1/8$.  In this case, the system no longer admits a solution in a form of a Fuchsian series, but instead there is a new family of solutions given by the form,
\begin{equation}
x=\frac{1\pm i\,\sqrt{3}}{A}t^{-1}.
\end{equation}
The scale factor here is an oscillatory function of time,  increasing or decreasing exponentially, depending on the value of $A$.
There is another interesting case describing  anti-decaying fluids  when $\delta\rightarrow \infty$, which in terms of energy transfer means that  $\rho_2$ decays while $\rho_1$ gains energy instead. From the third asymptotic decomposition of the system and when $B>0$, this oscillatory behaviour of solution is given by
\begin{equation}
x=\mp i\sqrt{2/B}\,t^{-1}+c_{41}t^3\mp i\frac{B}{12}\sqrt{2/B}c^2_{41}t^7+\cdots .
\end{equation}
Lastly, regarding the all-terms-dominant decomposition, there is a dynamical situation stemming from the first dominant balance asymptotically, that is  for $\delta=1/9$ there is a valid  series expansion given by, \begin{equation}
x=\frac{6}{A}t^{-1}+c_{11}t^{-2}+c_{21}t^{-3}+\cdots ,\end{equation}
and describes a universe driven away from any finite-time singularities.

It is interesting that many of these solutions exist in a generalized framework when there is curvature and  more general interactions, and this more general problem is currently under investigation in Ref. \cite{kittou2}. There appear to be  new regimes that exist only under the joined asymptotic influence of curvature and interaction between the fluids.

\end{document}